\documentclass[aps,prl,twocolumn,showpacs]{revtex4}

\usepackage{mathptm}
\usepackage{amsmath}    
\usepackage{dcolumn}                    
\usepackage{bm}                         
\usepackage{graphicx}

\newcommand {\ba} {\begin{eqnarray}}
\newcommand {\ea} {\end{eqnarray}}
\newcommand {\bq} {\begin{equation}}
\newcommand {\eq} {\end{equation}}

\usepackage{times}

\begin{document}

\title{The Absence of Renormalization of the Specific Heat Coefficient of the Interacting Fermion Systems}

\author{Yunkyu Bang}
\email[ ]{ykbang@chonnam.ac.kr} \affiliation{Department of
Physics, Chonnam National University, Kwangju 500-757, and Asia
Pacific Center for Theoretical Physics, Pohang 790-784, Korea}

\begin{abstract}
Contrary to the longtime and widely conceived belief, we proved
that the specific heat coefficient $\gamma$ --also called
Sommerfeld coefficient -- of the interacting Fermion system is not
renormalized by the wave-function renormalization factor $Z$ as
far as the system remains a Fermi liquid state.
\end{abstract}


\date{\today}
\maketitle

{\it Introduction --} Fermi liquid theory\cite{Landau, LW-I, L-II}
is the most fundamental conceptual building block of the modern
quantum theory of the interacting fermion systems such as metals,
semiconductors, superconductors, liquid $^3$He, neutron stars,
etc. In a nutshell, it suggests that an interacting fermion system
can be one-to-one mapped to a non-interacting fermion system for
the low energy excitations. In the process of this adiabatic
mapping, the essential effect of the interaction is to renormalize
the original bare fermions into a renormalized fermionic "{\it
quasi-particles}".  While the charge and spin quantum numbers --
when they exist -- of the fermions are protected by the gauge
invarience\cite{Ward} and relativity, respectively, the mass of
the fermion in the condensed matter is an effective mass from the
beginning and can be renormalized to be a different value from the
original effective mass $m_0$ of the non-interacting limit.
Therefore, the renormalized effective mass $m^{*}$ of the
quasiparticle (q.p.) is the single most important quantity which
determines the low energy properties of the interacting fermion
systems. Hence, the reliable measurement of this quantity by
experiments should be of principal importance to study the nature
and strength of the interaction of the fermionic system.

There are several different probes to measure the effective mass:
specific heat (SH) coefficient, de Haas-van Alphen (dHvA) effect,
angle resolved photo-emission spectroscopy (ARPES), optical
spectroscopy, etc. Although some interpretations might be
necessary to extract the value of $m^{*}$ from the above listed
measurements, theoretically all these measurements should provide
consistent information about the effective mass $m^{*}$. For
example, the ARPES measures the q.p. energy dispersion $E({\bf
k})$ $vs$ momentum ${\bf k}$ and directly shows us, without
interpretation, how heavy or light the q.p.s move. The dHvA effect
similarly depends on the q.p. dispersion $E({\bf k})$, so that its
measurement also provides a direct information of the renormalized
mass. However, since the construction of the Landau Fermi liquid
phenomenology\cite{Landau} and its theoretical
justifications\cite{LW-I, L-II, Nozieres, Pines Nozieres}, the
most commonly used probe for the effective mass of the q.p.s in
the Fermi liquid systems is the measurement of the SH coefficient.
In particular, Luttinger had shown in his seminal paper
\cite{L-II} in 1960 that the SH coefficient $\gamma$ ($\equiv
\lim_{T \rightarrow 0} C(T)/T$) should be enhanced from the
non-interacting value $\gamma_0$ such as $\gamma / \gamma_0 =
m^{*} /m_0$. Since then, the measurement of $\gamma$ has been
established as the most important tool to measure the effective
mass of the fermionic q.p.s. in the condensed matter systems.

In this paper, we showed that there was an error in the proof of
Luttinger and the SH coefficient $\gamma$ of the interacting
fermion system is not fully renormalized so that $\gamma /
\gamma_0 = m^{*} /m_0$ is not true. Our finding should have far
reaching consequences in the study of various interacting fermion
systems such as strongly correlated metals, liquid $^3$He, neutron
stars, etc. In this paper, we will be focusing only on the
questions of where was wrong in the Luttinger's proof and what is
the correct answer for the SH coefficient $\gamma$ of the
interacting fermion systems.

{\it SH coefficient $\gamma$ and DOS --} It is well known that the
SH coefficient of the non-interacting fermion system $\gamma_0$ is
given by \cite{AM}
\begin{equation}
\label{eq1}
\lim_{T \rightarrow 0} C(T)/T \equiv \gamma_0= \frac{\pi^2}{3}N_0(0),
\end{equation}
\noindent where $N_0(0)$ is the density of states (DOS) of the
non-interacting fermion system at the chemical potential.
Intuitively, the SH coefficient of the interacting fermion system
$\gamma$ is expected to be given with the above equation by
replacing $N_0(0)$ by the DOS of the interacting fermion system
$N(0)$ such as $\gamma= \frac{\pi^2}{3}N(0)$. But this absolutely
reasonable intuition falls in a serious trouble as follows. The
DOS $N(0)$ of the interacting fermion system can be calculated if
we know the exact one-particle Green's function which is formally
written as
$G(k,\omega)=\frac{1}{\omega-\epsilon(k)-\Sigma(k,\omega)}$ with
the exact self-energy $\Sigma(k,\omega)$.
However, we can show that $N(0)/N_0(0) \neq m^{*} /m_0$ and that
even the inequality $N(0)/N_0(0) >1$ is not guaranteed, as shown
below. This finding is in stark contrast to the common knowledge
that the SH coefficient should be enhanced by interaction such as
$\gamma / \gamma_0 \approx m^{*}/m_0 > 1$. There are two possible
options to resolve this dilemma: (1) $\gamma= \frac{\pi^2}{3}N(0)$
is not true for the interacting system; or (2) the common belief
$\gamma / \gamma_0 \approx m^{*}/m_0$ is wrong. The main
conclusion of this paper is that the option (2) is the correct
answer, namely, $\gamma$ does not measure the effective mass
$m^{*}$ of the renormalized fermionic q.p.s.

Let us begin with calculating $N(0)$. It is well known that the
self-energy in the Fermi liquid state has the well defined
expansion such as  $\lim _{T, \omega \rightarrow 0}
\Sigma(k,\omega) = \Sigma(k_F,0) + y_k \epsilon(k) -\lambda_k
\omega - i \delta$ \cite{LW-I,L-II}, where $y_k=\frac{\partial
\Sigma(k,0)}{\partial \epsilon}|_{k_F}$ and
$\lambda_k=\frac{\partial \Sigma(k_F,\omega)}{\partial
\omega}|_{\omega=0}$, respectively. Then
\begin{eqnarray}
\label{eq2}
N(0) &\equiv& -\frac{1}{\pi} \sum_k Im G(k,\omega=0) \\
&=& N_0(0) \lim _{\omega
\rightarrow 0}  \int \frac{d \epsilon}{\pi} ~~Im \frac{-1}{[1+\lambda_k] \omega -[1+y_k]\epsilon + i \delta} \\
& = & N_0(0) \lim _{\omega
\rightarrow 0}  \int \frac{d \epsilon}{Z_k}~~ \delta(\omega-\frac{Y_k}{Z_k}\epsilon) \\
&=& \frac{N_0(0)}{Y}
\end{eqnarray}

\noindent where the wave-function renormalization factor
$Z_k=1+\lambda_k ~( Z_k > 1)$ and the static renormalization
factor $Y_k=1+y_k$ are defined, respectively, and $Y=<Y_k>_{FS}$
the Fermi surface (FS) average of $Y_k$. The important point of
Eq.(5) is that the wave-function renormalization factor $Z_k$
--which is always larger than 1 due to the causality -- completely
drops in the exact DOS $N(0)$ of the interacting fermion system.
As can be seen in the $\delta-$function term of Eq.(4), the q.p.
dispersion is renormalized as $E(k)=\epsilon(k)\frac{Y_k}{Z_k}$ in
accord with the common knowledge. However, the reduction of the
q.p. spectral weight by $\frac{1}{Z_k}$ reduces the enhanced q.p.
DOS $N_{qp}(0)=N_0(0)\frac{Z}{Y}$ (where $Z=<Z_k>_{FS}$) into
$N_0(0)\frac{1}{Y}$ as shown in Eq.(5).

The exact DOS $N(0)=\frac{N_0(0)}{Y}$ is still renormalized by the
static renormalization factor $Y$. However, although there is no
general constraint to guarantee $Y >1$ or $Y <1$ as in the case of
$Z>1$, the known cases, such as the Hartree-Fock exchange
correction with the Coulomb potential, indicate that $Y
>1$ is usually satisfied \cite{HF} unless the Fermi liquid state becomes unstable.
This implies that the exact DOS defined in Eq.(2) tends to be
reduced by interaction, quite contrary to the common knowledge. In
this paper, however, we will mainly focus on the dynamic
renormalization factor $Z$, because $Z$ is the dominant
renormalization effect in most of the strongly interacting fermion
systems.

To demonstrate the correctness of the result of Eq.(5), we show
the numerical results of $N(\omega) = -\frac{1}{\pi} \sum_k Im
G(k,\omega)$ of a simple toy model in Fig.1(a) neglecting the
static renormalization effect (i.e. setting $Y=1$). In this
examplary calculations, we assumed a box like DOS for the
non-interacting fermion system as $N_0(\omega)=1.0$ for $-\Lambda
< \omega < \Lambda$ and the effect of interaction is simulated by
the Fermi liquid type self-energy $Im \Sigma(\omega)=\alpha
\omega^2$ for $-\Lambda < \omega < \Lambda$ including the
corresponding real part $Re \Sigma(\omega)$. We chose $\Lambda=5$.
The results are self-explaining, showing $N(0) =N_0(0)$ for all
interaction strength of $\alpha$. Increasing the interaction
strength, the width of the q.p. DOS around $\omega=0$ becomes
progressively narrowed and the spectral weight outside of it is
depleted toward the high energy region which is not fully
displayed here but the total spectral weight of the DOS should be
conserved. The width of the q.p. DOS around $\omega=0$ is roughly
proportional to $\sim 1/Z$ and the value of $Z$ is determined by
the combination of the interaction strength $\alpha$ and the band
width scale $\Lambda$.

\begin{figure}
\noindent
\includegraphics[width=90mm]{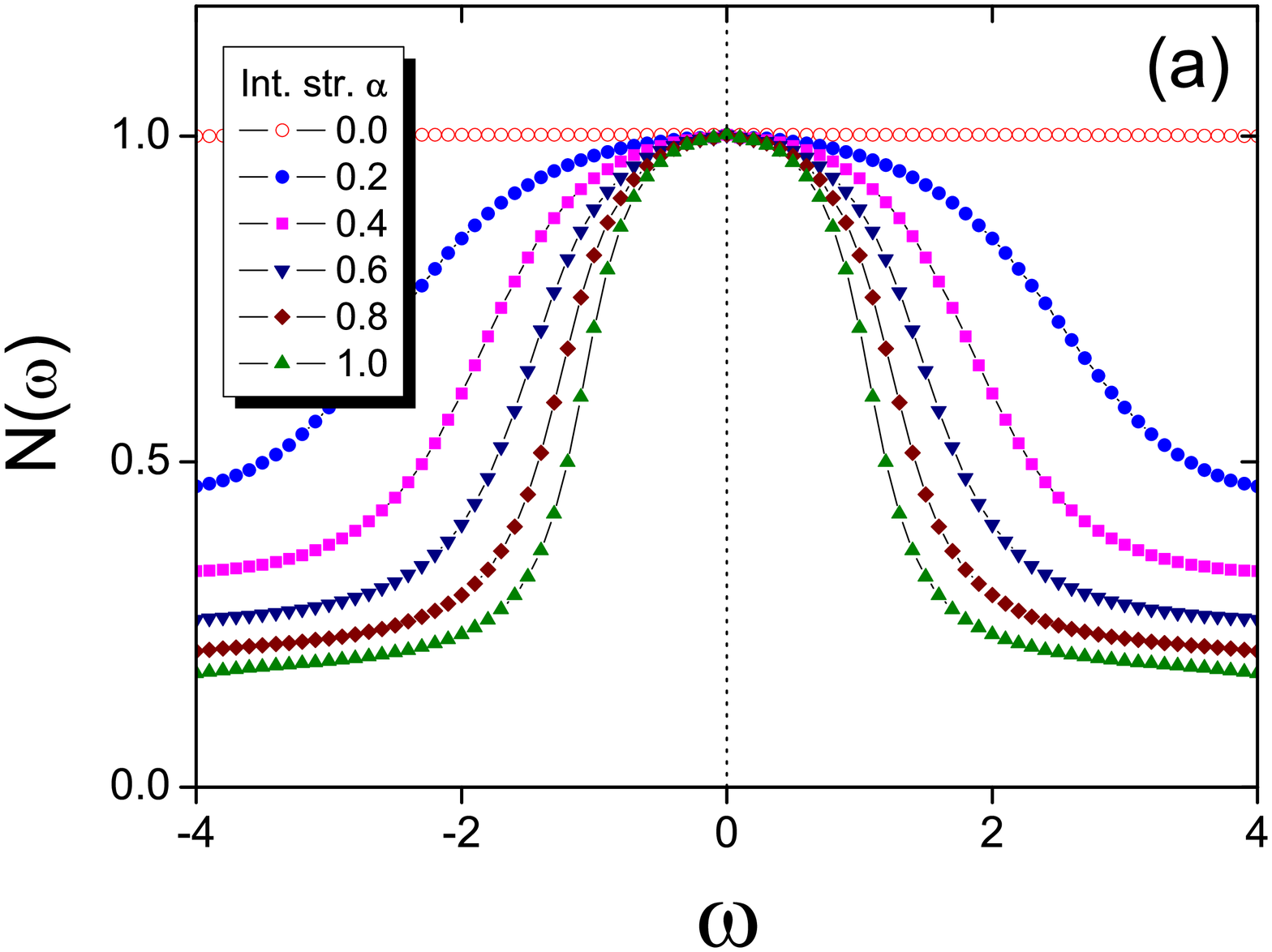}
\vspace{-0.5cm}
\includegraphics[width=90mm]{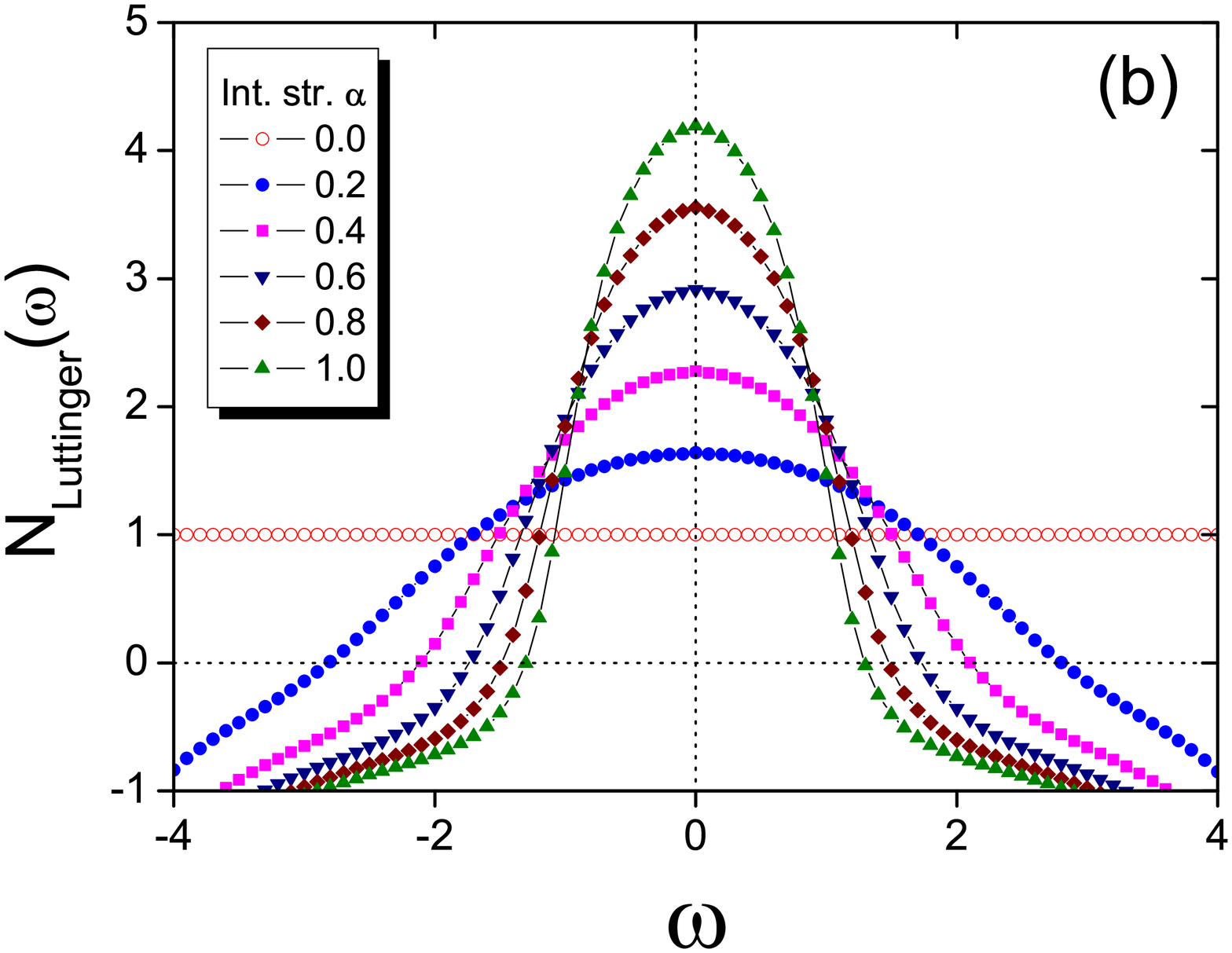}
\caption{(Color online) (a) The full DOS $N(\omega) = -\frac{1}{\pi} \sum_k Im G(k,\omega)$
with a phenomenological Fermi liquid type self-energy
$Im \Sigma(\omega)=\alpha \omega^2$ with $\alpha=0, 0.2, 0.4, 0.6, 0.8$ and 1.0, respectively
(the K-K related $Re \Sigma(\omega)$ is included).
A constant DOS $N_0 (\omega)=1.0$ was chosen for the non-interacting case ($\alpha=0$).
(b) The results of calculations of
$-\frac{1}{\pi} \sum_k Im \{ G(k,\omega) (1 - \frac{\partial \Sigma}{\partial
\omega}) \}$ with the same self-energies as in (a).
Note the relation $N_{Luttinger}(0)= Z \cdot N(0)$. \label{fig1}}
\end{figure}

{\it SH coefficient $\gamma$ of Interacting Fermi Systems  --} To
find an exact theoretic formula to calculate the SH coefficient
$\gamma$ of the interacting Fermi systems, we start with the same
Hamiltonian for the interacting fermion system used by Luttinger
and Ward \cite{LW-I}
\begin{eqnarray}
\label{Hamilton}
H = \sum_{r} \epsilon_r c^{\dag}_r c_r + \frac{1}{2} \sum_{r,s,r^{'},s^{'}}
c^{\dag}_r c^{\dag}_s c_{r^{'}} c_{s^{'}} (rs|v|r^{'} s^{'})
\end{eqnarray}
\noindent where $\epsilon_r$ is the energy measured from the
chemical potential of the non-interacting single particle states
with the index $r=(k,\sigma)$ for both momentum and spin.
$c^{\dag}_r, c_r$ are the creation and annihilation fermion
operators, respectively, and $(rs|v|r^{'} s^{'})$ is the general
four point fermion interaction matrix.

In Ref.\cite{LW-I}, Luttinger and Ward wrote down the celebrated
free energy functional of the interacting fermion system as

\ba \label{FEF} \Omega(T) &=& -T \sum_{r,n} e^{i \omega_n 0^{+}}
\{
\ln[\epsilon_r + \Sigma_r(\omega_n) -i \omega_n] \nonumber \\
& & + G_r (\omega_n)\Sigma_r(\omega_n)\} + \Omega^{'}(T)\ea

\noindent where $\omega_n =\pi T (2 n+1)$ is Matsubara frequency.
$G_r (\omega_n)$ and $\Sigma_r(\omega_n)$ are the full Green's
function and the full proper self-energy, respectively. The
functional $\Omega^{'}$ is defined by LW (refers to
Ref.\cite{LW-I}) as

\begin{equation}
\label{F'}
\Omega^{'} = \left(
\begin{array}{l}
\textrm{contribution of all closed-linked skeleton diagrams,}\\
\textrm{but with replacing all Green's function lines by} \\
\textrm{the full Green's functions}~~ G_r (\omega_n).
\end{array}
\right)
\end{equation}
The explicit expression of $\Omega^{'}$ was given in LW(50) (this
notation means Eq.(50) of Ref.\cite{LW-I}), but for our purpose we
don't need to know the details of the structure of $\Omega^{'}$.
The functional $\Omega^{'}$ was ingeniously designed by Luttinger
to satisfy the famous variational theorem of the total free energy
functional:
\bq \label{variation} \frac{\partial\Omega }{\partial \Sigma_r} =
0 .\eq
And this theorem can be satisfied only if the functional
$\Omega^{'}$ satisfies the following variational property
\bq \label{LW51} \frac{\partial\Omega^{'} }{\partial
\Sigma_r(\omega_n)} =T \sum_{r,n} [G_r(\omega_n)]^2
\Sigma_r(\omega_n) \eq

\noindent which was shown in LW(51). Up to now, we have just
copied the key results of  Ref.\cite{LW-I}. For our purpose, we
only need one slight generalization of Eq.(10) as follows
\bq \label{LW51'}\frac{\partial\Omega^{'} }{\partial i\omega_n} =
- T \sum_{r,n} [G_r(\omega_n)]^2 \Big(1- \frac{\partial
\Sigma_r}{\partial i\omega_n} \Big) \Sigma_r(\omega_n). \eq
The proof of Eq.(11) is easily deduced from Eq.(10) if we note the
expression of $G_r ^{-1}= i \omega_n - \epsilon_r -
\Sigma_r(\omega_n)$ and the trivial relations
\bq \frac{\partial G_r(\omega_n) }{\partial \Sigma_r(\omega_n)} =
[G_r(\omega_n)]^2 \eq and
\bq \frac{\partial G_r(\omega_n) }{\partial i\omega_n} = -
[G_r(\omega_n)]^2 \Big(1- \frac{\partial \Sigma_r}{\partial
i\omega_n}\Big).\eq
The Eq.(11) is the crucially important relation for our purpose
and will be used later.

In order to calculate the entropy from the free energy functional
Eq.(\ref{FEF}), we need to extract the leading temperature
dependent parts of it. Using a standard method of the Matsubara
frequency summation, Eq.(\ref{FEF}) is written as

\ba \label{FEF-T} \Omega(T) &=& \oint \frac{d z}{2 \pi i} n_F (z)~
\sum_{r} \{
\ln [\epsilon_r + \Sigma_r(z) -z] \nonumber \\
& & + G_r (z)\Sigma_r(z) \} - \oint \frac{d z}{2 \pi i} n_F (z)~
\Omega^{'} (z) \ea

\noindent where all Matsubara frequencies of Eq.(7) are
analytically transformed to complex numbers as $i \omega_n =z$ and
the functional $\Omega^{'} (z)$ is also understood as $\Omega^{'}
(i \omega_n \rightarrow z)$ after replacing the overall Matsubara
frequency summation $-T \sum_{n}$ of the original functional
$\Omega^{'} (i \omega_n)$ by the contour integral $\oint \frac{d
z}{2 \pi i} n_F (z)$ with the Fermi-Dirac distribution function
$n_F (z)$. Now it is clear that there are only two places which
contain the temperature dependence in the above free energy
functional Eq.(\ref{FEF-T}): $n_F(z)$ and $\Sigma_r(z)$. As
Luttinger argued \cite{L-II}, the leading temperature dependence
should come from the explicit summation of $i \omega_n$
(equivalently in $n_F(z)$) and the temperature variation of
"$\Sigma_r(T)-\Sigma_r(T=0)$" is a higher order and should be
neglected. Therefore, using $S(T)=-d \Omega(T)/dT$, we can write
down $S(T)$ as follows

\ba \label{S(T)} S(T) &=& \int_{-\infty}^{\infty} \frac{d
\omega}{\pi T} \omega \Big[\frac{\partial n_F (\omega)}{\partial
\omega} \Big] \nonumber
\\
& \cdot & \sum_{r} Im \{ \ln {G^{-1}_r (\omega)} + G_r
(\omega)\Sigma^{0} _r(\omega) \} \nonumber \\ & - &
\int_{-\infty}^{\infty} \frac{d \omega}{\pi T} \omega
\Big[\frac{\partial n_F (\omega)}{\partial \omega} \Big]~ Im
\Omega^{'} (\omega), \ea

\noindent where the contour path of $\oint$ is deformed along the
real frequency axis {\it \`{a} la} the appendix of
Ref.\cite{LW-I}; the $\omega$-integration for $[-\infty, \infty]$
should be carried infinitesimally above the real axis, i.e. for
$\omega+ i \eta$. $\Sigma^{0} _r(\omega)$ means
$\Sigma_r(\omega,T=0)$ and it is understood that every
$\Sigma_r(\omega,T)$ implicit in the above expression is replaced
by $\Sigma^{0} _r(\omega)$. While the above expression of $S(T)$
is undoubtedly the exact expression, Luttinger argued in
Ref.\cite{L-II} that the leading temperature dependence of
$\Omega(T)$ (Eq.(\ref{FEF})) is contained only in

\bq \label{eq14} \Omega_{Luttinger}(T) \approx -T \sum_{r,n} e^{i
\omega_n 0^{+}} \ln[\epsilon_r + \Sigma_r(\omega_n) -i \omega_n]
\eq

\noindent and ignored the last two terms of Eq.(\ref{FEF}) because
the leading temperature dependent parts in the remaining terms $-T
\sum_{r,n} e^{i \omega_n 0^{+}} [G_r (\omega_n)\Sigma_r(\omega_n)]
+ \Omega^{'}$ cancels each other. Hence, Luttinger has obtained
the entropy from Eq.(16) as follows

\ba \label{S_Lutt} S_{Luttinger}(T) & = & \int_{-\infty}^{\infty}
\frac{d \omega}{\pi T} \omega \Big[\frac{\partial n_F
(\omega)}{\partial \omega} \Big] \sum_{r} Im \{ \ln {G^{-1}_r
(\omega)}\} \\
&=& \int_{-\infty}^{\infty} \frac{d \omega}{\pi T} \omega
\Big[\frac{\partial n_F (\omega)}{\partial \omega} \Big] \sum_{r}
Im \{ \ln [\epsilon_r + \Sigma^{0}_r -\omega] \}. \nonumber
\\ & & \ea
The above expression $S_{Luttinger}(T)$ is the only the first term
of the exact entropy expression $S(T)$ of Eq.(15). Then it is
obvious question how to justify using $S_{Luttinger}(T)$ to
calculate the SH instead of using the exact $S(T)$. The only
justification is that both expressions Eq.(15) and Eq.(17) give
the same result or put in other words the contributions of the
last two terms of Eq.(15) cancel each other as Luttinger claimed.
However, below we show that the cancellation between the two terms
are incomplete and an important contribution remains. Therefore we
have to use the full expression of the entropy Eq.(\ref{S(T)}).

Expectedly the calculation results of the SH coefficient $\gamma$
from $S(T)$ and $\gamma_{Luttinger}$ from $S_{Luttinger}(T)$ are
totally different: the former one yields $\gamma$ unrenormalized
by the wave-function renormalization factor $Z$ regardless of the
strength of the interaction while the latter one yields an
enhanced $\gamma_{Luttinger}$ proportional to the value of $Z$ as
widely believed in the community ever since the proof of Luttinger
\cite{L-II}.

To calculate $\gamma  \equiv \lim_{T \rightarrow 0} C(T)/T =
\lim_{T \rightarrow 0} d S(T)/d T$, we only need to extract
$T$-linear contributions in $S(T)$ or $S_{Luttinger}(T)$.
Utilizing Sommerfeld expansion, we then only need to extract
$\omega$-linear terms in the integrand of $Im{ ...}$ in $S(T)$ or
$S_{Luttinger}(T)$. Let us first calculate $\gamma_{Luttinger}$
from $S_{Luttinger}$. The leading Taylor expansion of the
integrand of $S_{Luttinger}$ can be read from Eq.(18) as

\ba \label{eq13} Im \{ \ln [\epsilon_r + \Sigma^{0}_r -\omega] \}
&=& Im G_r(\omega=0) (1 - \frac{\partial Re\Sigma_r ^{0}}{\partial
\omega}\Big|_{\omega=0}) \cdot \omega \nonumber \\
&+& O(\omega^2) ... \ea

\noindent  Using the notation $(1 - \frac{\partial Re\Sigma_r
^{0}}{\partial \omega}\Big|_{\omega=0})=1+\lambda_r = Z_r$ and
combining it with Eqs.(2)-(5), we obtained
\ba N_{Luttinger}(0) &\equiv& -\frac{1}{\pi} \sum_r Im
G_r(\omega=0) (1+\lambda_r) \\
&=&  N(0)\cdot Z = N_0(0)\cdot \frac{Z}{Y}\ea
\noindent where $Z$ is a Fermi surface average of $Z_r$. In
Fig.1(b), we showed the numerical calculations of $-\frac{1}{\pi}
\sum_r Im \big\{ G_r(\omega) (1 - \frac{\partial \Sigma_r
^{0}}{\partial \omega}) \big\}$ with the varying interaction
strength. This quantity has no direct physical meaning (it becomes
even negative at higher energies) but its zero frequency value
$N_{Luttinger}(\omega=0)$ clearly demonstrated the result of
Eq.(21) and showed what quantity was used by Luttinger for the
calculation of the SH coefficient.
Substituting the results of Eqs.(19)-(21) into Eq.(18), the
leading temperature dependence of $S_{Luttinger}$ is the following
\ba S_{Luttinger}(T) & \approx & - \int_{-\infty}^{\infty} \frac{d
\omega}{T} \omega \Big[\frac{\partial n_F (\omega)}{\partial
\omega} \Big] N_{Luttinger}(0) \cdot \omega \ea
\noindent and from this we can derive the same result as Luttinger
had obtained\cite{L-II} as
\bq \label{eq15} \gamma_{Luttinger} = \frac{\pi^2}{3} \cdot
N_{Luttinger}(0) = \frac{\pi^2}{3} N_0(0) \cdot \frac{Z}{Y},  \eq
\noindent so that the SH coefficient $\gamma_{Luttinger}$ is
indeed enhanced by the factor $Z$ compared to the non-interacting
case. Note that $N_{Luttinger}(0)$ defined in Eq.(20) is nothing
but the quasi-particle DOS $N_{qp}(0)$ which was conventionally
defined by re-scaling fermion operators $c_r$ by the factor
$\sqrt{Z_r}$. Hence the Luttinger's result of Eq.(23) has firmly
established that the SH coefficient $\gamma$ measures the q.p. DOS
$N_{qp}(0)$.

Now let us use the exact expression $S(T)$ of Eq.(15) to derive
$\gamma$. The coefficients of the $\omega$-linear terms of the
integrand of $S(T)$, $\{ \ln {G^{-1}_r (\omega)} + G_r
(\omega)\Sigma^{0} _r(\omega) \} - \Omega^{'}(\omega)$, are the
following
\ba \label{eq16} &=& G_r (1 - \frac{\partial \Sigma_r ^{0}}{\partial \omega}) \nonumber \\
&-& [G_r]^2 \Sigma^0 _r (1 - \frac{\partial \Sigma_r
^{0}}{\partial \omega}) + G_r \frac{\partial \Sigma_r
^{0}}{\partial \omega} \nonumber \\
&+& [G_r]^2 \Sigma_r^0 (1 - \frac{\partial \Sigma_r ^{0}}{\partial
\omega}) \nonumber \\
&=& G_r. \ea
\noindent Above we have arranged the Taylor expansions of each
three terms $ \ln {G^{-1}_r (\omega)}$, $G_r
(\omega)\Sigma^{0}_r(\omega)$ and $ - \Omega^{'}(\omega)$ into
three separate lines for clarity. In particular, we have used the
important relation of Eq.(11) for $\frac{\partial
\Omega^{'}}{\partial \omega}$ in the third line.
There are lots of cancellations and the final result should be
compared to Eq.(19) obtained from $S_{Luttinger}(T)$. In fact, the
above cancellation is the consistent result of the Luttinger's
variational theorem of Eq.(9) which requires that all variations
of $\partial \Sigma_r$ in the total free energy functional
$\Omega$ should sum up to zero \cite{LW-I}. In this sense, the
expression of $S_{Luttinger}$ in Eq.(18) with Eq.(19) cannot be
correct since it contains $\partial \Sigma_r$ term.

Now it is a trivial matter to calculate the SH coefficient
$\gamma$ substituting the result of Eq.(24) into Eq.(15) as
\ba S(T) & \approx & \int_{-\infty}^{\infty} \frac{d \omega}{\pi
T} \omega \Big[\frac{\partial n_F (\omega)}{\partial \omega} \Big]
\sum_r Im G_r(\omega=0) \cdot \omega  \nonumber \\
&=& - \int_{-\infty}^{\infty} \frac{d \omega}{T} \omega
\Big[\frac{\partial n_F (\omega)}{\partial \omega} \Big] N(0)
\cdot \omega\ea
\noindent and combining with Eqs.(2)-(5), we have \ba \label{eq26}
\gamma &=& \frac{\pi^2}{3} \cdot N(0) = \frac{\pi^2}{3} \cdot
\frac{N_0(0)}{Y}. \ea
The above result shows that the SH coefficient $\gamma$ of the
interacting Fermi system measures the exact DOS $N(0)$ defined by
Eq.(2), which is consistent with our physical intuition. However,
due to the absence of the wave-function renormalization factor $Z$
in contrast to the Luttinger's result of Eq.(23), we do not expect
a strong enhancement of $\gamma$ by the interaction in a Fermi
liquid state unless the static renormalization factor $Y$ becomes
$0 <Y <1$.

{\it Other Physical Quantities --} The renormalized q.p. mass
$m^{*} \approx m_0 \cdot \frac{Z}{Y}$ due to interaction is
measured by different experimental probes. Indeed the energy
dispersion of the q.p. pole $E(k)$, defined by $\omega -
\epsilon(k) - \Sigma(k, \omega) =0$, is renormalized as $E(k)
\approx \epsilon(k) \cdot \frac{Y}{Z}$ and should be directly
measured by ARPES without any interpretation or confusion. Another
common tool to measure $m^{*}$ is the dHvA effect with the applied
external field $H$. In this case, the effective mass is measured
from the temperature reduction factor of the signal strength which
is given by the Lifshitz-Kosevich formula $R_T \sim
\exp{(-T/\omega_c)}$ \cite{Schoenberg}, where $\omega_c$ is the
cyclotron frequency. $\omega_c$ is determined by the q.p. energy
distance between the Landau levels quantized by the field $H$ as
$\Delta E =\hbar \omega_c$, and the Landau level is determined by
the q.p. dispersion $E(k)$ to the first approximation, hence
$\omega_c=eH/m^{*}c$. Therefore, the dHvA effect measurement can
provide an information of $m^{*}$.
Lastly, the optical spectroscopy measurements need a more careful
interpretation. The total spectral density near the Fermi level is
not enhanced by interaction as shown in Eq.(5), but the width of
the q.p. dispersion is narrowed by the factor $1/Z$ as shown in
Fig.1(a). Therefore, for example, the width of the Drude spectra
in the optical conductivity is expected to be reduced by the
factor $1/Z$ ,while the absolute magnitude of the zero frequency
conductivity $\sigma(\omega=0)$ nor the total Drude spectral
weight is not expected to be enhanced. However, because the
optical conductivity is a transport property, it is essential also
to count on the renormalized Fermi velocity $\tilde{v}_{F}$ and
the scattering rate $1/\tau_{tr}$ due to the interaction besides
the q.p. DOS. Therefore, for more complete details of the optical
properties of the interacting fermion systems, we need to analyze
the two particle correlation function which is beyond the scope of
the current paper.

{\it Conclusions --} In summary, we have shown the following: (1)
Luttinger's calculation of $\gamma_{Luttinger}$ is not correct
because it started with an approximate functional
$\Omega_{Luttinger}$; (2) the SH coefficient $\gamma$ measures the
exact DOS $N(0)$ defined in Eq.(2) and is not enhanced by $Z$ the
wave-function renormalization factor; therefore, (3) the q.p. DOS
$N_{qp}(0)$ is only a fictitious concept and not a measurable
quantity.
These results are in stark contrast to the longtime accepted idea
of the interaction-enhanced SH coefficient since the proof of
Luttinger in 1960 \cite{L-II}. The implications of our finding
should be far reaching because the enhanced SH coefficient
$\gamma$ in the interacting Fermion systems has been accepted and
utilized for the last 50 years as a pivotal building concept in
the study of the interacting Fermi liquid systems both in theory
and in experiment. We need to rethink many of the previous ideas
and measurements based on this --{\it now proven wrong} --
concept.

{\it Acknowledgement -- } This work was supported by Grants No.
NRF-2010-0009523 and No. NRF-2011-0017079 funded by the National
Research Foundation of Korea.

\end{document}